\documentclass[preprint,aps,prl,floatfix,showpacs,showkeys]{revtex4}

\usepackage{graphicx,amsmath,amssymb,dcolumn}
\usepackage{bm}

\newcommand{\be}{$\beta$}

\newcommand{\su}[1]{$^{#1}$}
\newcommand{\msu}[1]{^{#1}}

\newcommand{\sub}[2]{$\emph{#1}_{\emph{#2}}$}
\newcommand{\msub}[2]{{#1}_{#2}}
\newcommand{\g}{$\gamma$}

\newcommand{\et}{\emph{et al.}~}
\newcommand{\degree}{$\,^{\circ}$}
\newcommand{\tdef}{$\msub{\bar\theta}{def}$}

\preprint{}

\begin{document}

\title{Nuclear spin polarization following intermediate-energy heavy-ion reactions}

\author{D.E. Groh,\su{1,2} J.S. Pinter,\su{1,2,} P.F. Mantica,\su{1,2} T.J. Mertzimekis,\su{2,}\footnote{Present address: Department of Physics, The University of Ioannina, Ioannina 45110, Greece} A.E. Stuchbery,\su{3} and D.T. Khoa\su{4}}
\affiliation{
$^{1}$Department of Chemistry, Michigan State University, East Lansing, MI 48824 USA\\
$^{2}$National Superconducting Cyclotron Laboratory, Michigan State University, East Lansing, MI 48824 USA\\
$^{3}$Department of Nuclear Physics, Research School for Physical Sciences and Engineering, The Australian National University, Canberra, ACT 0200, Australia\\
$^{4}$Institute for Nuclear Science and Technique, VAEC, PO Box 5T-160, Nghia Do, Hanoi, Vietnam}

\date{\today}
\begin{abstract}

Intermediate-energy heavy-ion collisions can produce a spin polarization of the projectile-like species.  Spin polarization has been observed for both nucleon removal and nucleon pickup processes.  Qualitative agreement with measured spin polarization as a function of the momentum of the projectile-like fragment is found in a kinematical model that considers conservation of linear and angular momentum and assumes peripheral interactions between the fast projectile and target.  Better quantitative agreement was reached by including more realistic angular distributions, de-orientation caused by \g-ray emission, and by correcting for the out-of-plane acceptance.  The newly introduced corrections were found to apply to both nucleon removal and nucleon pickup processes.

\end{abstract}


\maketitle

\section{Introduction}
Nuclear spin polarization is a necessary condition for many types of physics experiments including studies of nuclear structure, nuclear reactions, fundamental interactions, and condensed matter physics.  Many of these experiments require the polarization of radioactive ion beams (RIBs). For example, ground state nuclear magnetic dipole ($\mu$) and electric quadrupole moments ($Q$) are critical in the study of nuclear structure, especially as one moves away from the valley of stability.  Currently, the best reach for such measurements has been demonstrated by RIBs polarized in fragmentation reactions.  

Quenched magnetic moment values are observed for the extremely neutron-deficient nuclei \su{9}C and \su{57}Cu.  The small $\mu$ value of the proton drip line nucleus \su{9}C ($\msub{S}{p}$=1.3 MeV) leads to a large deviation of the deduced isoscaler expectation value $\langle\sigma\rangle$ for the \su{9}Li-\su{9}C mirror pair when compared to observed trends for $T=1/2$ nuclei \cite{Mat}.  Shell model calculations were shown to better reproduce $\langle\sigma\rangle$ when isospin mixing was included \cite{Huh}. Intruder mixing with the ground state of \su{9}C caused by proton shell quenching may also account for the deviant $\mu$ value \cite{Uts}.  \su{57}Cu represents the heaviest $\msub{T}{z}=-1/2$ nuclide with a known $\mu$ value.  The magnetic moment is smaller than predicted in shell model calculations that assume \su{56}Ni is a good doubly-magic nucleus \cite{KMin}.

The quadrupole moments of the B isotopes also provide evidence for changes in nuclear structure.  $Q(\msu{8}B)$ was found to be twice as large as expected from shell model calculations \cite{TMin}.  The larger $Q$ was considered evidence of a proton halo covering a neutron core.  Near the neutron drip line, the experimental $Q$ moments for odd-mass \su{13}B, \su{15}B, and \su{17}B were found to be nearly constant, independent of neutron number \cite{Oga}.  The calculated $Q$ obtained with standard values of effective charges overpredict the experimental $Q$ for \su{15}B and \su{17}B isotopes.  The result reveals that both proton and neutron quadrupole polarization charges may be reduced near the neutron drip line \cite{Sag}.

From these examples, it is apparent that moment measurements play an important role in nuclear structure studies.  As stated earlier, in order to perform any moment measurement, the nuclei must be spin polarized.  There is a strong dependence of polarization on the momentum of the fragment nucleus, therefore, it is crucial to know the magnitude of polarization prior to the experiment.  While fragmentation reactions are key in producing exotic nuclei, these nuclei come with low rates.  A figure of merit for measurements involving polarization is $P\msu{2}Y$, where $P$ represents polarization and $Y$ is yield.  The optimization of polarization with yield is important.  Improvements in yield will come with the development of new RIB facilities, but while yields are small, the ability to accurately predict the expected polarization is required for experimental success.  Broader applications for the polarized beams will be achieved with new RIB facilities, but at the same time, better predictive power is needed.

In general, the spin orientation of an ensemble of quantum states is usually specified by a statistical tensor \cite{Ald}.  The tensor can be denoted by $\msub{\rho}{kq}(I)$, where $I$ is the spin of the state and $k$, the rank of the statistical tensor, takes the values $k=0,1,2,...2I$ if $I$ is an integer, and $k=0,1,2,...2I-1$ if $I$ is half-integer.  $q$ takes integer values between $-k$ and $+k$.  The statistical tensor is related to the distribution of magnetic substates with respect to a chosen coordinate frame.  It is usually possible to specify the magnetic substate distribution by the populations $P(m)$ of the $2I+1$ $m$-substates, where it is assumed that the $P(m)$ are normalized so that $\sum P(m)=1$.

For an $m$-state distribution with axial symmetry, only the $q=0$ components of $\msub{\rho}{kq}(I)$ are non-zero.  Spin polarization is defined in terms of the $k=1$ statistical tensor, denoted $\msub{\rho}{10}$, relative to its value for maximum polarization.  Specifically,
\begin{equation}
\msub{\rho}{10}(I)=-\sum_{m}\frac{mP(m)}{\sqrt{I(I+1)}}
\end{equation}
and
\begin{equation}
\msub{\rho}{10}\msu{max}(I)=\frac{-I}{\sqrt{I(I+1)}}
\end{equation}
so the spin polarization is
\begin{equation}
\frac{\msub{\rho}{10}(I)}{\msub{\rho}{10}\msu{max}(I)}=\sum_{m}\frac{mP(m)}{I}\equiv\left\langle\frac{\msub{I}{z}}{I}\right\rangle
\end{equation}
Thus, spin polarization is a measure of the orientation of the total angular momentum relative to a fixed axis. 

\begin{figure}[hbp!]
\includegraphics[width=7cm]{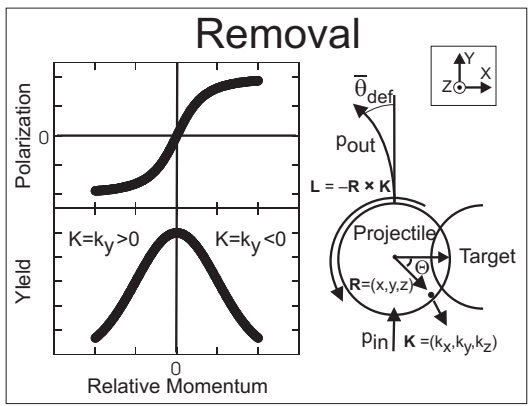}
\caption{Illustration of nuclear spin polarization produced in a nucleon removal reaction at intermediate energies, for a high $Z$ target.  The yield and polarization curves are given relative to the incident projectile momentum.  The removal schematic is given in the projectile-like rest frame.} 
\label{fig:excelsimRemoval}
\end{figure}

Spin polarization of projectile-like residues from intermediate-energy heavy-ion reactions was first observed at RIKEN \cite{Asa}.  Fragments detected at small angles with respect to the normal beam axis were shown to have a polarized spin in the low-intensity wings of the momentum distribution.  A qualitative description of the polarization mechanism was found in a model that considers conservation of linear and angular momentum, and assumes peripheral interactions between the fast projectile and target. Fig.\ \ref{fig:excelsimRemoval} presents a schematic of the expected polarization and yield for the nucleon removal process for fragmentation of a projectile on a heavy target. A systematic study of the spin polarization following few-nucleon removal from light projectiles as a function of energy and target was completed by Okuno \et \cite{Oku}.  This study demonstrated that the relation between the outgoing fragment momentum and sign of spin polarization depended on the mean deflection angle, \tdef. Near-side reactions occur for high-$Z$ targets, where the Coulomb deflection will dominate the internuclear potential between projectile and target, giving the polarization dependence shown in Fig.\ \ref{fig:excelsimRemoval}.  The nucleon-nucleon potential energy governs removal reactions on low-$Z$ targets. Far-side reactions prevail in this case, and the sign of the observed polarization is reversed.

The spin polarization has a near-zero value at the peak of the fragment yield curve for both near- and far-side dominated reactions, since $\mid$\tdef$\mid$ is large. This behavior can be qualitively understood from the projectile rest-frame diagram in Fig.\ \ref{fig:excelsimRemoval}. The removed nucleons have momentum \textbf{K}. The $z$-component of the induced angular momentum of the projectile-like species is $\msub{\ell}{z}=-X\msub{k}{y}+Y\msub{k}{x}$, where $X, Y$ are the localized cartesian coordinates of the removed nucleon(s), and $\msub{k}{x}, \msub{k}{y}$ are the momentum components of the removed nucleons in the reaction plane. If the nucleon removal occurs uniformly in the overlap region, $X\sim\msub{R}{0}, Y\sim0$, then $\msub{\ell}{z}=-X\msub{k}{y}$.  Zero polarization will therefore result when the fragment momentum equals the projectile momentum, since $\msub{k}{y}=0$ in the projectile rest frame under these conditions.

If nucleon removal is not uniform in the overlap region, $Y\ne0$ and the term $Y\msub{k}{x}$ can contribute to $\msub{\ell}{z}$.  Such a contribution will only be observed experimentally when $\mid$\tdef$\mid$ is small.  The final scattering angle of the fragment is $\msub{\theta}{L}=$\tdef$+\Delta\theta$, where $\Delta\theta$ is the change in angle caused by the transverse momentum component of the removed nucleons, $\Delta\theta=\tan\msu{-1}(-\msub{k}{x}/p)$.  Here $p$ is the total momentum of the projectile-like fragment.  In reactions where $\mid$\tdef$\mid\sim0$, it is the transverse momentum component of the removed nucleon(s) that ``kicks'' the fragments to small angles, and the resulting polarization is negative since $\msub{k}{x}>0$ to give positive $\Delta\theta$ and $Y<0$ for non-uniform nucleon removal as illustrated in Fig.\ \ref{fig:excelsimRemoval}.

A Monte Carlo code was developed \cite{Oku} based on the ideas discussed above to simulate the spin polarization generated in nucleon removal reactions at intermediate energies. The general behavior of the spin polarization as a function of projectile-like momentum was achieved, although a scaling factor of 0.25 was needed to reproduce the magnitude of polarization observed experimentally.

\begin{figure}[hbp!]
\includegraphics[width=7cm]{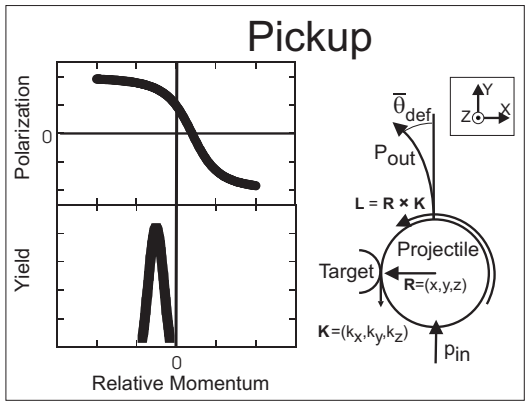}
\caption{Illustration of nuclear spin polarization produced in a nucleon pickup reaction at intermediate energies.  The yield and polarization curves are given relative to the incident projectile momentum.  The pickup schematic is given in the projectile-like rest frame.} 
\label{fig:excelsimPickup}
\end{figure}

Spin polarization via nucleon pickup reactions at intermediate energies was first demonstrated at NSCL \cite{Gro}.  Positive spin polarization was determined for \su{37}K species collected at small angles in the reaction of \su{36}Ar projectiles on a \su{9}Be target at 150 MeV$/$nucleon. Fig.\ \ref{fig:excelsimPickup} illustrates the spin polarization and yield from nucleon pickup reactions. The key to understanding the observed spin polarization in the pickup process is that the picked-up nucleon must have an average momentum equal to the Fermi momentum oriented parallel to the beam direction. Souliotis \et \cite{Sou} reached this conclusion from the observed shifts in the centroids of the momentum distributions for one and two-nucleon pickup products. The average projectile-like momentum $<p>$ was found to satisfy the relation $<p>=<\msub{p}{p}>+<\msub{p}{t}>$, with $<\msub{p}{p}>$ the average momentum of the incident particle and $<\msub{p}{t}>$ the average momentum of the picked up nucleon equal to the Fermi momentum.  In the rest frame of the projectile-like species, the momentum of the picked-up nucleon will be antiparallel to the incoming projectile momentum.

The $z$-component of orbital angular momentum induced by the nucleon pickup process is $\msub{\ell}{z}=R\Delta p$, assuming a peripheral interaction where the nucleon is picked up to a localized position on the projectile given by $R$ in Fig.\ \ref{fig:excelsimPickup}. The spin polarization will be zero when the momentum of the picked-up nucleon matches the momentum of the incoming projectile ($\Delta p=0$).  The zero crossing occurs at the projectile-like momentum $p=[(\msub{A}{p}+1)/\msub{A}{p}]\msub{p}{p}$, where $\msub{A}{p}$ and $\msub{p}{p}$ are the mass number and momentum of the projectile, respectively.  A linear increase in $\msub{\ell}{z}$ is expected with a decrease in the momentum of the outgoing pickup product. Groh \et \cite{Gro} found that proton pickup reactions follow the trend shown in Fig.\ \ref{fig:excelsimPickup}, except for the low momentum side of the momentum distribution.  At low momentum values of the pickup products, the matching conditions for pickup will no longer be satisfied, and the spin polarization is observed to rapidly approach zero.

Turz\'o \et showed that neutron pickup reactions at intermediate energies behave in a similar manner \cite{Tur}.  They extended the Monte Carlo simulation of Ref. \cite{Oku} to include nucleon pickup and the momentum considerations discussed by Groh \et \cite{Gro}. Qualitative agreement of the observed spin polarization as a function of the projectile-like product was realized, as was the case with nucleon removal reactions. But again, a scaling factor of 0.25 was needed to reproduce the magnitude of the observed spin polarization. Scaling factors of the same magnitude required for both nucleon removal and nucleon pickup suggest that the same quantitative correction factors should apply to both.

The kinematical model proposed by Asahi \et has been successfully employed to
qualitatively explain spin polarization at intermediate energies for both nucleon removal and pickup reactions. This paper introduces additional considerations to the kinematical model aimed to improve the quantitative agreement of the Monte Carlo simulation with the experimentally-observed spin polarization in nucleon removal and pickup reactions.

\section{Monte Carlo Simulation}
 
The simulation code developed by Okuno \et \cite{Oku} was used to compute the momenta of the removed nucleons in Monte Carlo fashion. The kinematical equations from Ref. \cite{Asa} were then applied to calculate the spin polarization $(\msub{\ell}{z}/L)$ as a function of fragment momentum.  The nucleon removal positions $X, Y$ on the projectile surface are determined by the relations $X=\msub{R}{0}\cos\Theta$ and $Y=-\msub{R}{0}\sin\Theta$, where $\Theta>0$ as shown in Fig.\ \ref{fig:excelsimRemoval}. Improvements made to the original simulation code, which advance toward better quantitative agreement with spin polarization in both nucleon removal and nucleon pickup reactions, are discussed below.

\subsection{Location of Nucleon Abrasion}
\begin{figure}
\includegraphics[width=7cm]{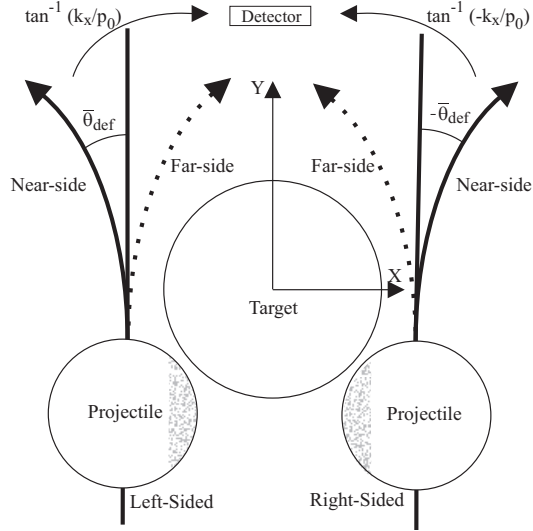}
\caption{Schematic diagram of right- and left-sided collisions.  The dotted lines are far-side interactions and the solid lines are near-side interactions.} 
\label{fig:sidedness}
\end{figure}

One shortcoming of the original simulation code was that the projectile-target interaction was one-sided. We have adopted an absolute coordinate system to maintain the correct relationship between the sign of the polarization, the emission angle of the fragment, and the momentum of the fragment.  Interactions between projectile and target are permitted in the reaction plane, on both ``sides" of the target, as shown in Fig.\ \ref{fig:sidedness}. The coordinate system is defined as follows.  Positive $y$ is the beam direction, positive $x$ is defined to the right of the target relative to the beam direction, and positive $z$ is perpendicular to the scattering plane forming a right-handed coordinate system.  Positive angles are defined to the left of the $y$-axis, or toward negative $x$.  For left-sided interactions, a near-side(far-side) collision will scatter to the left(right), or to positive(negative) angles.  For the interactions on the right side of the target, a near(far)-side interaction must scatter to the defined negative(positive) angles, thus the signs of the mean deflection angles are changed for these events.  

\begin{figure}
\includegraphics[width=7cm]{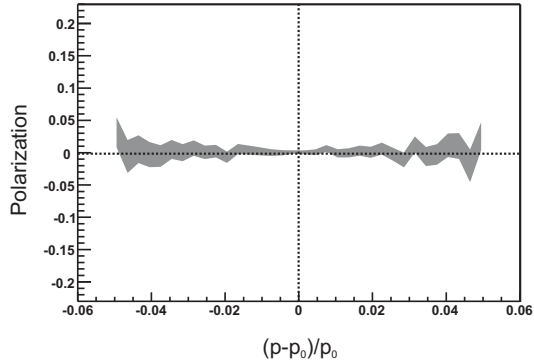}
\caption{(Color online) Calculated polarization as a function of relative fragment momentum for the fragmentation of 109.6 MeV$/$nucleon \su{15}N in a target of \su{197}Au to make \su{13}B at a fragment angle of 0\degree.  \sub{p}{0} is the momentum of the incident projectile and $p$ is the momentum of the outgoing fragment.  The grey band represents the range of simulation results within a $1\sigma$ distribution.} 
\label{fig:zerodeg}
\end{figure}

One of the fundamental phenomena in fragmentation reactions related to polarization is that no fragment spin polarization is observed when the mean fragment angle, $\msub{\theta}{L}$, is zero degrees \cite{Man}.  Shown in Fig.\ \ref{fig:zerodeg} is the calculated polarization for 109.6 MeV$/$nucleon \su{15}N fragmented in a target of \su{197}Au to make \su{13}B at a mean fragment angle of 0\degree{} with an angular acceptance of 0.25\degree.  The calculated polarization is equivalent to zero within statistical error.  Allowance for right- or left-sided collisions correctly accounts for the observed absence of spin polarization.  On average, half of the events detected at 0\degree{} are from a right-sided interaction, while the other half are from left-sided interactions.  The resulting polarization for each type of interaction is equal in magnitude, but opposite in sign, giving an average of zero polarization.  

Previously, an average value for the location $X, Y$ of the removed nucleons in the projectile was calculated based on the rotation angle $\Theta$. The same removal location was then used for every event in the simulation.  The removal location is now calculated based on the volume of intersection of a cylinder and sphere following the prescription of Gosset \et \cite{Gos}. The number of nucleons removed is directly proportional to the overlap volume. A removal position was calculated in Monte Carlo fashion for each nucleon in the overlap region and then averaged to give the position $X,Y$ of the group of removed nucleons. The rotation angle $\Theta$ is accommodated in this approach by offseting the $Y$ position by a fixed value, yet ensuring that the offset $Y$ value remains in the overlap region.

\subsection{Distribution of Deflection Angles}
The mean deflection angle \tdef{}  is calculated by way of numerical integration. Required input parameters are the $Z$ and $A$ of both projectile and target, the energy of the projectile, the distance of closest approach, $\msub{r}{min}$, and the real part of the optical model potential for the nucleus-nucleus interaction, $\msub{V}{0}$. The relations of Gossett \et \cite{Gos} are used to determine $\msub{r}{min}$.

In the original incarnation of the simulation code, all resulting fragments scatter to a single angle defined by \tdef, based on the number of nucleons removed plus the angular impulse given to the fragment in the transverse direction.  However, for a given number of nucleons removed, fragments will scatter to a range of angles whose mean is the mean deflection angle.  To better simulate experimental conditions, a distribution of scattering angles whose average is \tdef{} has been considered, including Rutherford, Fermi, and step function distributions. The Rutherford distribution describes classical elastic scattering and is strongly peaked at 0\degree.  This distribution was problematic because fragmentation reactions are not classical elastic scattering, and the Rutherford distribution is incompatible with a mean angle larger than about 1\degree.  The step function distribution ranged in angle from zero to twice the mean deflection angle, with the functional form shown below:
\begin{equation}
f(\msub{\theta}{def})=\left\{ \begin{array}{ll}
1 & 0 \leq \msub{\theta}{def} < 2\msub{\bar\theta}{def}\\
0 & \msub{\theta}{def} \geq 2\msub{\bar\theta}{def}
\end{array} \right.
\end{equation}
Even though the mean is \tdef, the angular distribution is non-physical.  In particular, the maximum number of fragments did not occur at 0\degree{} as observed in experiment.  The Fermi distribution (i.e. a Woods-Saxon shape) has the form
\begin{equation}
f(\msub{\theta}{def})=\frac{1}{1+e\msu{(\msub{\theta}{def}-x)/a}}
\end{equation}
where $a=0.5$, $\msub{\theta}{def}$ is the scattering angle and the value of $x$ is varied to make the mean of the distribution equal to the mean deflection angle.  The correct mean for the distribution could only be achieved with a negative value for $x$, which is unphysical.  

\begin{figure}[hbp!]
\includegraphics[width=7cm]{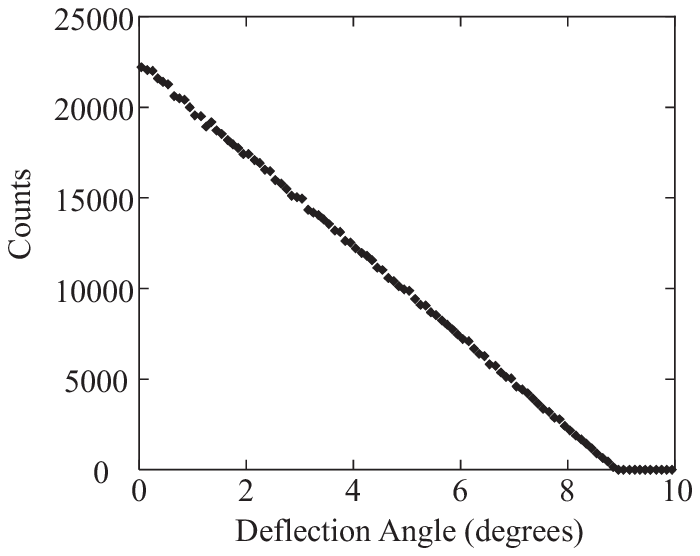}
\caption{Linear angular distribution with \tdef=3\degree.} 
\label{fig:defangle}
\end{figure}

A compromise was made on a straight-line distribution with negative slope.   This distribution has the advantage of peaking at 0\degree, and by changing the slope, the desired mean deflection angle can be imposed on the distribution. A calculated angular distribution with \tdef=3\degree{} is shown in Fig.\ \ref{fig:defangle}.

\subsection{Out-of-Plane Scattering}
\begin{figure}[hbp!]
\includegraphics[width=7cm]{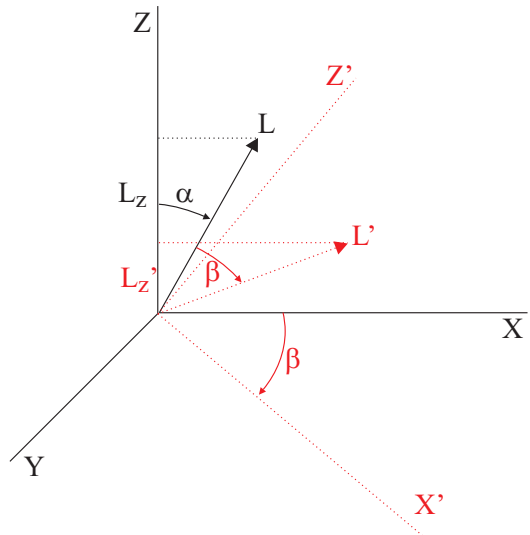}
\caption{(Color online) Representation of non-equatorial scattering of fragments (primed axes) and the scattering of fragments purely in the horizontal $x-y$ plane (unprimed axes).  Labels defined in the text.} 
\label{fig:outofplane}
\end{figure}

In the original simulation code, projectile-target interactions that occurred only in the $x, y$ plane were considered. However, fragments should interact with the target on the top, the bottom, and all the angles in between, not just on the right or left side of the target in the horizontal x-y plane. Such non-equatorial interactions would decrease the $z$-component of angular momentum in the equatorial (reaction) plane, as shown in Fig.\ \ref{fig:outofplane}.  The primed frame represents non-equatorial scattering and can be represented as a rotation of the coordinate system about the y-axis through the angle \be.  

The resulting $z$-component of angular momentum for the non-equatorial scattering is $\msub{L}{z}\msu{'}=\msub{L}{z}\cos\beta-\msub{L}{x}\sin\beta$, where $\msub{L}{x}$ is the component of angular momentum projected on the $x$ axis and \be{}  is the rotation angle.  In theory, \be{} can range from $-\pi/2$ to $+\pi/2$, since the projectile interaction can occur anywhere in the $yz$ plane.  Integrating all possible contributions from the out-of-plane acceptance, and including a $1/\pi$ normalization factor from the interval of integration gives 
\begin{equation} \label{eq:Lz}
\msub{L}{z}\msu{'}=\frac{1}{\pi}\int_{-\frac{\pi}{2}}^{\frac{\pi}{2}}(\msub{L}{z}\cos\beta-\msub{L}{x}\sin\beta)d\beta=\frac{2}{\pi}\msub{L}{z}
\end{equation}

\begin{figure}[hbp!]
\includegraphics[width=7cm]{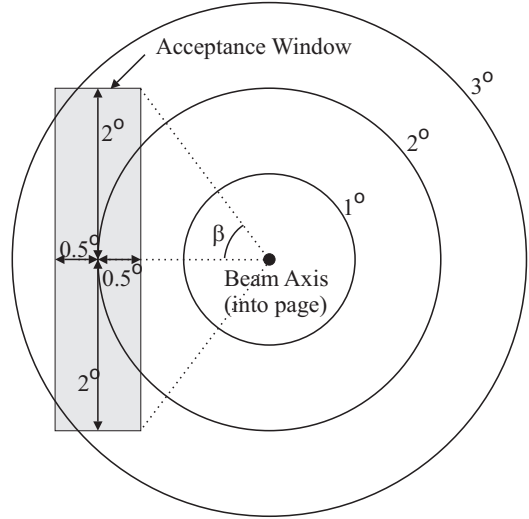}
\caption{Representation of the angular acceptance window from a beam view.  The fragment angle acceptance (horizontal acceptance) is 2.0\degree$\pm$0.5\degree{} and the vertical acceptance is $\pm$2\degree.} 
\label{fig:outofplane2}
\end{figure}

Experimental devices have a limited angular acceptance, and thus the range on \be{} will be smaller than $-\pi/2$ to $+\pi/2$.  Fig.\ \ref{fig:outofplane2} shows a schematic of the beam view of an angular acceptance of 2.0\degree$\pm$0.5\degree{} horizontally and $\pm$2\degree{} vertically.  
The out-of-plane acceptance of $\pm$2\degree{} limits the range on \be{} to -53\degree{} to +53\degree{} for this case.  Equation \ref{eq:Lz} with these limits of integration and a re-normalization of 106\degree{} rather than 180\degree{} gives $\msub{L}{z}\msu{'}=0.86\msub{L}{z}$.  As the acceptance window moves further from the beam axis, the range on the angle \be{} will decrease.  This causes the value of $\msub{L}{z}\msu{'}$ to approach that of $\msub{L}{z}$.  Various example acceptance windows, \be{} angles (i.e. ranges on \be) and the resulting corrections for $\msub{L}{z}$ in terms of $\msub{L}{z}\msu{'}$ are listed in Table \ref{tab:outofplane}.  A vertical acceptance of $\pm$2\degree{} is assumed in all cases, common in most fragment separators.

\begin{table}[hbp!]
\caption{Polarization correction factors due to non-equatorial scattering.}
\label{tab:outofplane}
\begin{tabular}{|c|c|c|c|}
\hline
\textbf{Fragment angle}(\degree)   &   \textbf{Acceptance}(\degree)   &   $\msub{\beta}{max}$(\degree)   &   \textbf{Correction factor} \\
\hline
1.0 & $\pm$2.5 & 90 & 0.64 \\
2.0 & $\pm$0.5 & 53 & 0.86 \\
2.0 & $\pm$2.5 & 90 & 0.64 \\
4.0 & $\pm$2.5 & 53 & 0.86 \\
5.0 & $\pm$2.5 & 39 & 0.92 \\
0   & full     & 90 & 0.64 \\
\hline
\end{tabular}
\end{table}

A factor of $2/\pi$ has been included in the simulation code as a multiplicative factor on the polarization.  As shown in Table \ref{tab:outofplane}, different angular acceptance windows could slightly increase this value.

\subsection{\g-ray Deorientation}

The fragmentation process will often leave the projectile-like product in an excited state. One pathway to remove the excitation energy is through \g-ray emission. Discrete \g{} rays have been observed in ``in-beam'' fragmentation studies \cite{Yon}. Intermediate-energy reactions also populate high-spin isomeric states, whose depopulation can be readily followed in the absence of the prompt radiation background \cite{Grz}. \g{} rays emitted from a spin polarized nucleus can reduce the magnitude of the polarization. Such a deorientation process was not previously considered.  \g-ray deorientation was included in the simulation code by assuming a statistical cascade through a continuum of levels following a prescription similar to that of Leander \cite{Lea}. A nuclear level density is specified by a constant temperature level-density formula \cite{von}. For each \g{} ray, the transition energy, multipolarity ($E1, E2,$ or $M1$), and spin change are determined by random numbers. The deorientation coefficients $\msub{U}{k}$ \cite{Ham} are calculated from the initial and final spin values, and the multipolarity. A cumulative deorientation is obtained from the product of the $\msub{U}{1}$ values of the cascade \g{} rays. The statistical nature of the cascade decay is included by repeating the calculation with a random walk. The resulting average deorientation coefficients are given a weighting factor based on the calculated spin distribution and excitation energy of the fragment.

Deorientation coefficients were calculated for the fragmentation reactions involving two-nucleon removal reported in Ref.\ \cite{Oku}, as well as the nucleon pickup reactions reported in Refs.\ \cite{Sou} and \cite{Tur}.  The values are presented in Table \ref{tab:gray}. The $\msub{U}{1}$ values ranged from 0.50 to 0.89, suggesting that \g-ray deorientation can have significant impact on the final spin polarization of the projectile-like fragment. The calculated average entry spin ($\sim2\hbar$) and cascade multiplicity ($\sim3$) were reasonable based on in-beam spectroscopy of light fragmentation products \cite{Yon}. 

\begin{table}[hbp!]
\caption{Polarization correction factors due to \g-ray deorientation}
\label{tab:gray}
\begin{tabular}{|c|c|c|c|c|}
\hline
\textbf{Reaction}   &   \textbf{Energy (MeV$/$A)}   &   $\msub{U}{1}$   &   \textbf{Avg. Entry Spin} & \textbf{Multiplicity} \\
\hline
\su{197}Au(\su{14}N,\su{12}B)X & 39.4 & 0.500 & 1.54 & 2.32  \\
\su{197}Au(\su{15}N,\su{13}B)X & 68 & 0.592 & 1.79  & 2.35\\
\su{197}Au(\su{15}N,\su{13}B)X & 109.6 & 0.576 & 1.71  & 2.34\\
\su{93}Nb(\su{15}N,\su{13}B)X & 67.3 & 0.583 & 1.74  & 2.34\\
\su{27}Al(\su{15}N,\su{13}B)X & 68  &  0.555 & 2.02 & 2.39 \\
\su{9}Be(\su{36}Ar,\su{37}K)X & 150.0 &  0.688 & 2.59 & 2.95  \\
\su{9}Be(\su{36}S,\su{34}Al)X & 77.6 &  0.887 & 4.17 & 2.82  \\
\hline
\end{tabular}
\end{table}

It can be anticipated that nuclear structure effects, especially the discrete level sequence at low energy, may affect the magnitude of deorientation in some cases.  For example, \su{11}Be has an $I\msu{\pi}=1/2\msu{+}$ ground-state and its only excited state at 320 keV has $I\msu{\pi}=1/2\msu{-}$.  Any population of the ground-state via the excited state will have the opposite polarization to the direct population of the ground-state.  This is an exceptional case.  In $p$-shell nuclei closer to the valley of stability that have low lying 1/2\su{+} and 1/2\su{-} states, with one being the ground state, the excited spin 1/2 state is accompanied by several other states of higher spin that decay directly to the ground state.  Thus, if the population of the excited states is not selective, the effect of the change in the sign of the polarization for decays through excited spin 1/2 states is not expected to be prominent in most cases.

Low-lying spin 0 states also have the potential to greatly diminsh the net polarization.  For example, in \su{16}N, which has a 2\su{-} ground state, the first-excited state at 120 keV is a 0\su{-} state.  Any population that passes through this state will lose its orientation.  However, in \su{16}N there are 3\su{-} and 1\su{-} states at 298 and 397 keV, respectively, which will compete with the 0\su{-} state for population as a result of the reaction, and ameliorate the overall polarization loss.

Thus although specific nuclear level sequences at low spin are important, especially in weakly bound systems with very few excited states, as a first estimate of the magnitude of the effect it seems appropriate to model the deorientation in a generic way by assuming a statistical cascade through a continuum of levels using a Monte-Carlo simulation as described above.

\subsection{Nucleon Pickup}
\begin{figure}
\includegraphics[width=7cm]{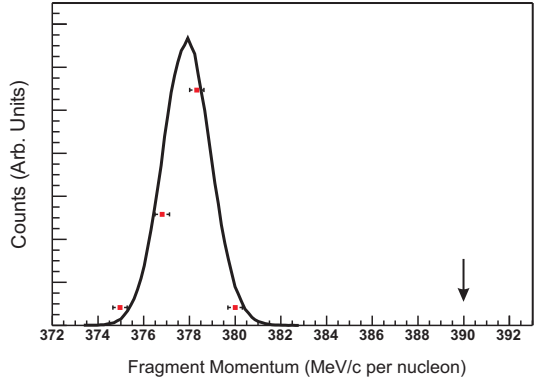}
\caption{(Color online) Parallel momentum$/$nucleon distribution calculated with the simulation code for the reaction \su{27}Al(\su{18}O,\su{19}O) at 80 MeV$/$nucleon.  The red squares are the data points and the black line represents the simulation results.  The arrow corresponds to the momentum/nucleon of the beam.  The simulated momentum distribution has been scaled by the ratio observed in Ref.\ \cite{Sou} of experimental centroid$/$calculated centroid (0.969$/$0.978), in order to compare to the data.}
\label{fig:momentum}
\end{figure}

The simulation code has been modified to include nucleon pickup, independent of the efforts reported in Ref. \cite{Tur}.  The pickup process follows the observations of Souliotis \et \cite{Sou}, in that the picked-up nucleon has an average momentum equal to the Fermi momentum (230 MeV$/$c), oriented parallel to the beam direction.  The momentum distribution for the one-neutron pickup reaction \su{27}Al(\su{18}O,\su{19}O) at 80 MeV$/$nucleon shows a considerable shift of the centroid below the momentum$/$nucleon of the beam, as observed in Ref.\ \cite{Sou}, in contrast to those for nucleon removal products.   The simulated position of the centroid agrees with the calculation of Ref.\ \cite{Sou}, where a simple model based on momentum conservation was used (see Fig.\ \ref{fig:momentum}).  

The width of the momentum distribution is observed experimentally to be small (around 20 MeV$/$c), while it is calculated to be zero, based on the equation by Goldhaber \cite{Gol}, and extended to nucleon pickup:
\begin{equation}
\label{eq:Goldhaber}
\msub{\sigma}{\Vert}\msu{2}=\msub{\sigma}{0}\msu{2}\frac{\msub{A}{PF}(\msub{A}{P}-\msub{A}{PF})}{\msub{A}{P}-1}
\end{equation}
where $\msub{A}{PF}=\msub{A}{F}-\Delta\msub{A}{t}$ is the mass of the projectile part of the final product and $\Delta\msub{A}{t}$ is the number of nucleons picked up from the target.  The parameter $\msub{\sigma}{0}$ is the reduced width, and is related to the Fermi momentum of the nucleon motion inside the projectile:  $\msub{\sigma}{0}\msu{2}=\msub{p}{Fermi}\msu{2}/5$.  A value of $\msub{\sigma}{0}=80$ MeV$/$c is used, which agrees with the experimental widths in Ref.\ \cite{Sou}.  Eq.\ \ref{eq:Goldhaber} assumes that the nucleon is picked up from the target with a fixed momentum and direction, and the picked up nucleon makes no contribution to the width.  Thus, for any pure nucleon pickup process, $\msub{A}{P}=\msub{A}{PF}$ and the parallel width is zero. In order to model experimental observations of Ref.\ \cite{Sou}, a parallel width of 20 MeV$/$c was used.  In addition to the parallel width, Van Bibber \et \cite{Van} showed that in heavy-fragment studies in the 100 MeV$/$nucleon region, the projectile is subject to an orbital deflection due to its interaction with the target nucleus before fragmentation takes place.  The orbital deflection gives an additional dispersion of the transverse momentum, as demonstrated in the expression:
\begin{equation}
\msub{\sigma}{\bot}\msu{2}=\msub{\sigma}{1}\msu{2}\frac{\msub{A}{PF}(\msub{A}{P}-\msub{A}{PF})}{\msub{A}{P}-1}+\msub{\sigma}{2}\msu{2}\frac{\msub{A}{PF}(\msub{A}{PF}-1)}{\msub{A}{P}(\msub{A}{P}-1)}
\end{equation}
where the first term was defined previously, and the second term contains $\msub{\sigma}{2}\msu{2}$, the variance of the transverse momentum of the projectile at the time of fragmentation (200 MeV$/$c as used in Ref.\ \cite{Van}).  A comparison of the simulated momentum distribution is shown in Fig.\ \ref{fig:momentum} with the data taken from Ref.\ \cite{Sou}.

\subsection{Optical Potential}
\begin{figure}
\includegraphics[width=7cm]{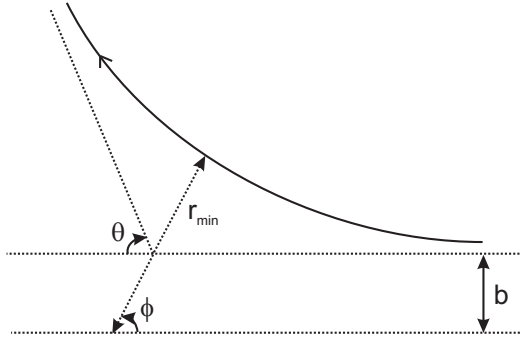}
\caption{Variable definitions for mean deflection angle calculation.}
\label{fig:meandef}
\end{figure}

The real part of the optical model potential for the nucleus-nucleus interaction, \sub{V}{0}, is a required input parameter for the mean deflection angle calculation.  For a single interaction, the deflection angle $\theta$ (see Fig.\ \ref{fig:meandef}) is given by   
\begin{equation}
\theta=\pi-2\phi
\end{equation}
with 
\begin{equation}
\phi=\int_{\msub{r}{min}}^{\infty}\frac{b dr}{r\msu{2}\sqrt{1-\frac{b\msu{2}}{r\msu{2}}-\frac{U(r)}{E}}}
\end{equation}
where $b$ is the impact parameter, $r$ is the distance between the centers of the two objects, $U(r)$ is the potential governing the interaction of the two objects, $\msub{r}{min}$ is the separation between the centers of the two point-like objects at the distance of closest approach and the energy, $E$, is given by
\begin{equation}
E=\frac{1}{2}m\msub{v}{\infty}\msu{2}
\end{equation}
where $\msub{v}{\infty}$ is the velocity of the projectile at $r=\infty$ \cite{Lan}.

The projectile is assumed to move away from the target after the scattering event with momentum equal to the incident momentum, thus $E(\msub{v}{\infty})=E(\msub{v}{incident})$.  This formula is general for any spherically symmetric potential.

The potential $U(r)$ is defined by
\begin{equation}
U(r)=\msub{U}{Coulomb}(r)+\msub{U}{nuclear}(r).
\end{equation}
The Coulomb part of the potential is repulsive and is equal to
\begin{equation}
\msub{U}{Coulomb}(r)=\frac{1.438\msub{Z}{p}\msub{Z}{t}}{r}
\end{equation}
where $\msub{Z}{p}$ and $\msub{Z}{t}$ are the charges on the projectile and target respectively, and $r$ is the separation in fm.  The nuclear part of the potential is based on the real part of the optical model \cite{Kra}, and is attractive:
\begin{equation}
\msub{U}{nuclear}(r)=\frac{-\msub{V}{0}}{1+e\msu{(r-R)/a}}.
\end{equation}
Here $\msub{V}{0}$ is the depth of the optical model potential, $R=1.2(\sqrt[3]{\msub{A}{p}}+\sqrt[3]{\msub{A}{t}})$ where \sub{A}{p} and \sub{A}{t} are the masses of the projectile and target respectively, and $a$ is a measure of the diffuseness of the nuclear surface.  \sub{V}{0} and $a$ are parameters fit to experimental data.  There are very limited nucleus-nucleus scattering data, and an exact determination or parametrization of \sub{V}{0} is difficult for any given projectile-target combination.  Typically this is not a problem because in head-on collisions, the nuclear potential does not have a large influence.  However, the treatment of peripheral collisions depends on the optical potential.  In the minimum, a determination of \sub{V}{0} is needed.  A parametrization of \sub{V}{0} based on energy and/or number of nucleons removed would suffice, but unfortunately, such a parametrization does not presently exist.  

In the reactions studied in Ref.\ \cite{Oku}, \sub{V}{0} was calculated by the authors, and the corresponding \tdef{} was reported in the literature.  The reported \tdef{} (given in Table \ref{tab:V0}) was used in the polarization calculations to follow.  

For the nucleon pickup reactions, \sub{V}{0} was determined with a folding model calculation \cite{Kho}.  The model was chosen because it reproduces experimental scattering data for heavy ions in the energy range of interest.  The folding calculation yields the real part of the optical potential (\sub{V}{0}) as a function of the internuclear radius, the distance between the center of the projectile and target. The internuclear radius is calculated in the simulation code, based on the relations by Gosset \et \cite{Gos}, as mentioned previously.  The value of \sub{V}{0} corresponding to a radius for one nucleon overlap was used in the nucleon pickup reactions.  Given in Table \ref{tab:V0} are the \tdef{} and \sub{V}{0} for the reactions studied in the present work.

\begin{table}[hbp!]
\caption{\tdef{} and corresponding \sub{V}{0} used for the reactions studied in the present work.}
\label{tab:V0}
\begin{tabular}{|c|c|c|c|}
\hline
\textbf{Reaction}   &   \textbf{Energy (MeV$/$A)}   &   \tdef(\degree)   &   \sub{V}{0} (MeV)  \\
\hline
\su{197}Au(\su{14}N,\su{12}B)X & 39.4 & 3.09 & 48   \\
\su{197}Au(\su{15}N,\su{13}B)X & 68.0 & 0.92 & 63  \\
\su{197}Au(\su{15}N,\su{13}B)X & 109.6 & 0.09 & 85  \\
\su{93}Nb(\su{15}N,\su{13}B)X & 67.3 & -0.35 & 65  \\
\su{27}Al(\su{15}N,\su{13}B)X & 68.0  &  -3.25 & 118 \\
\su{9}Be(\su{36}Ar,\su{37}K)X & 150.0 &  -0.07 & 29  \\
\su{9}Be(\su{36}S,\su{34}Al)X & 77.6 & -0.49 & 32 \\
\hline
\end{tabular}
\end{table}

\subsection{Results}
\begin{figure}
\includegraphics[width=13cm]{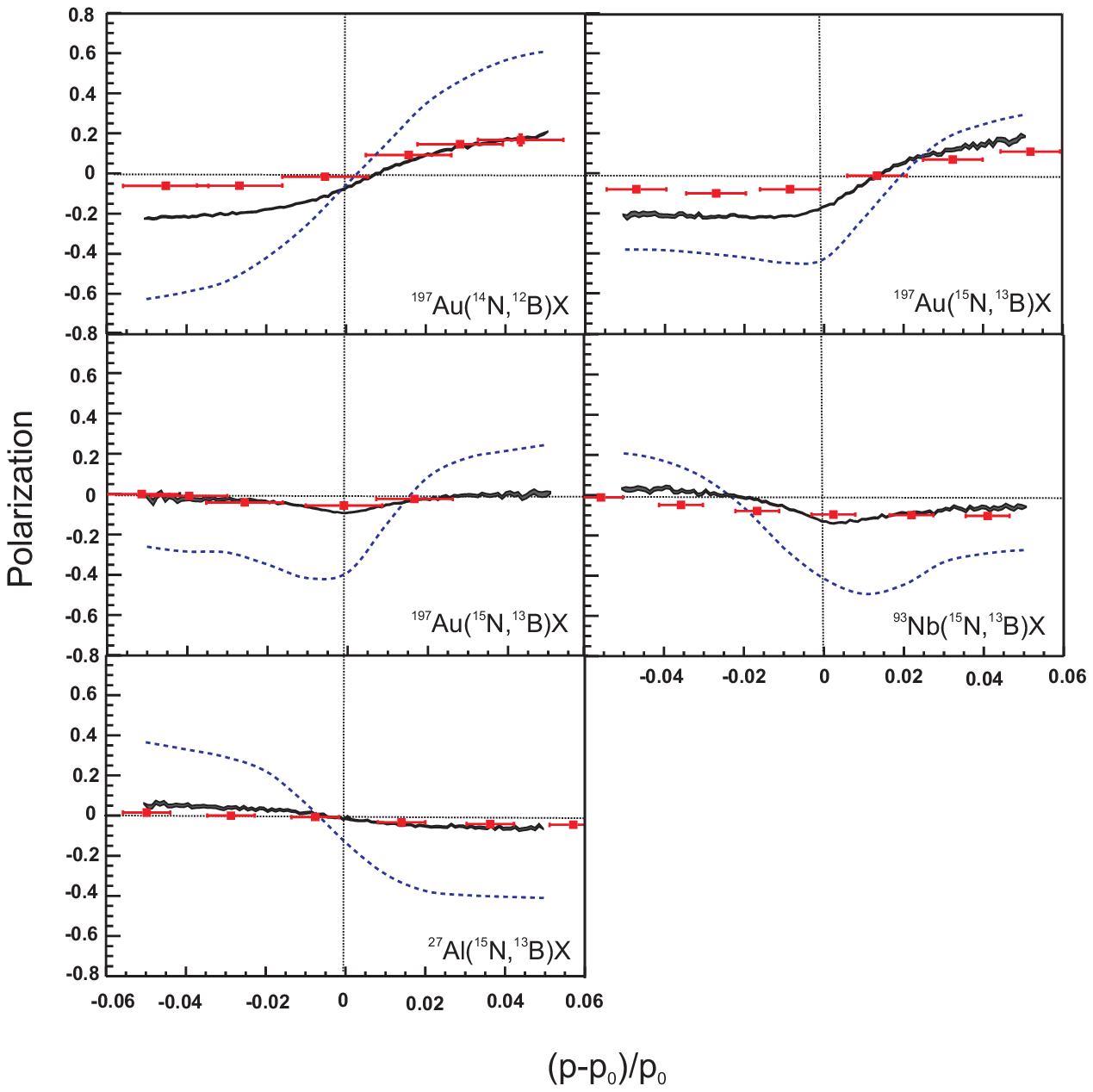}
\caption{(Color online) Polarization as a function of fragment momentum $p$ relative to the primary beam momentum $\msub{p}{0}$ for the removal reactions \su{197}Au(\su{14}N,\su{12}B) at 39.4 MeV$/$nucleon, \su{197}Au(\su{15}N,\su{13}B) at 68 MeV$/$nucleon, \su{197}Au(\su{15}N,\su{13}B) at 109.6 MeV$/$nucleon, \su{93}Nb(\su{15}N,\su{13}B) at 67.3 MeV$/$nucleon, and \su{27}Al(\su{15}N,\su{13}B) at 68 MeV$/$nucleon.  The red squares are the experimental data points and the blue dashed lines are the previous simulation results, both from Ref.\ \cite{Oku}.  The black band represents the range of the present simulation results within a $1\sigma$ distribution.  Momentum is given relative to the peak of the yield distribution.} 
\label{fig:RemovalDataSim}
\end{figure}

Comparison of the spin polarization measurements for intermediate-energy reactions that include two-nucleon removal in Ref. \cite{Oku} with the simulation results that include considerations for angular distributions, out-of-plane scattering, and \g-ray deorientation, are presented in Fig.\ \ref{fig:RemovalDataSim}. No scaling factor was used to adjust the simulated spin polarization.

\begin{figure}
\includegraphics[width=15cm]{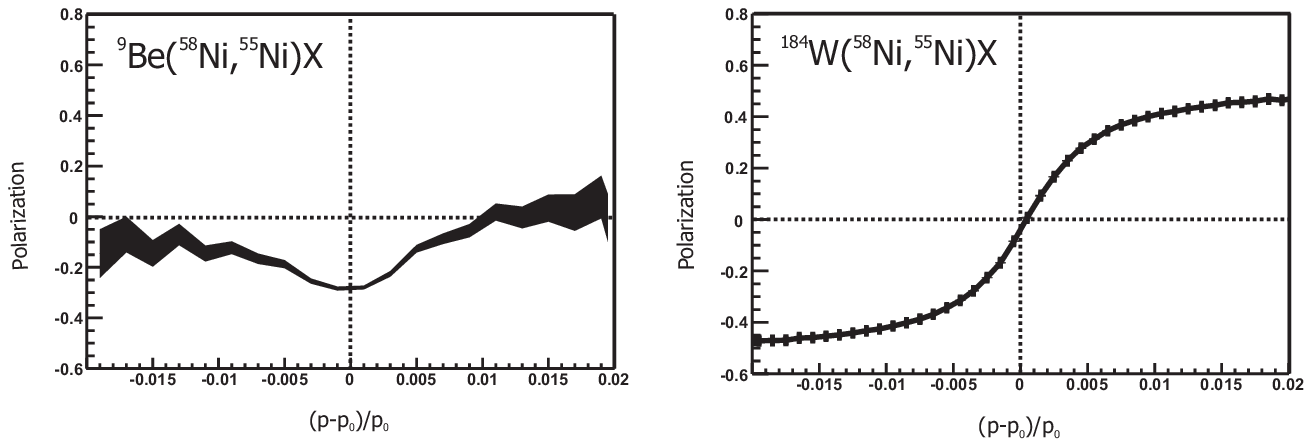}
\caption{Polarization distributions generated from the Monte Carlo simulation code as a function of relative fragment momentum for the three-neutron removal fragmentation reaction with a primary \su{58}Ni beam at 140 MeV$/$nucleon on a a$)$ \su{9}Be target and b$)$ \su{184}W target.  Both targets will be used during the experiment, to provide better comparison to the simulation.} 
\label{fig:55Ni}
\end{figure}

The predicted polarization trend as a function of outgoing fragment momentum for a three-nucleon removal case is shown in Figure \ref{fig:55Ni}. 
The polarized \su{55}Ni nucleus from a \su{58}Ni primary beam is proposed to determine the ground state magnetic moment measurement of \su{55}Ni at NSCL \cite{Exp}.  The expected rate of \su{55}Ni is 500 pps, thus optimized polarization at peak yield is critical to realize a successful measurement.  The proposed measurement will complete the $\mu$ measurements of nuclei one nucleon removed from \su{56}Ni.  The two $\msub{T}{z}=+1/2$ nuclei, \su{55}Co and \su{57}Ni, both have known magnetic moment values that compare favorably with results of shell model calculations.  On the other side of $N=Z$, the $\msub{T}{z}=-1/2$ nucleus \su{57}Cu has a $\mu$ value that deviates significantly from shell model expectations, suggesting a weakened core in \su{56}Ni.  The measurement of $\mu$(\su{55}Ni) is key to understanding how the \su{56}Ni core may be polarized when coupled to a neutron hole.  

\begin{figure}[hbp!]
\includegraphics[width=8cm]{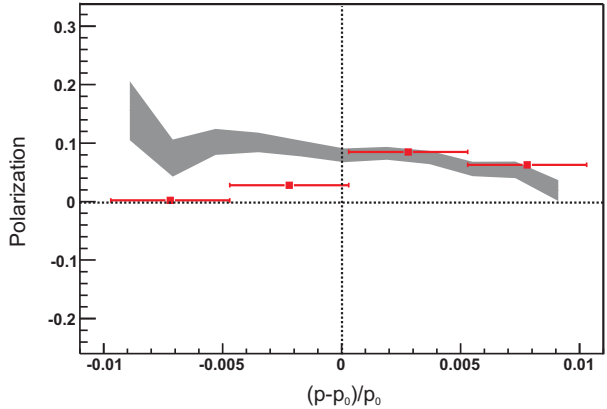}
\caption{(Color online) Polarization as a function of fragment momentum $p$ relative to the primary beam momentum $\msub{p}{0}$ for the one-proton pickup reaction \su{9}Be(\su{36}Ar,\su{37}K) (150 MeV$/$nucleon).  The red squares are the experimental data points from Ref. \cite{Gro} and the grey band represents the range of simulation results within a $1\sigma$ distribution.}
\label{fig:37Kband}
\end{figure}

\begin{figure}
\includegraphics[width=8cm]{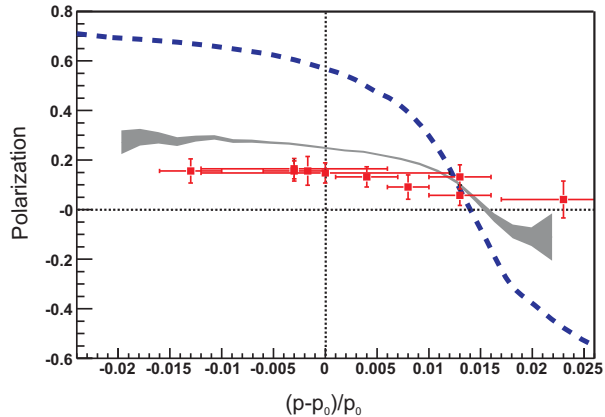}
\caption{(Color online) Polarization as a function of fragment momentum $p$ relative to the momentum at the peak of the yield distribution $\msub{p}{0}$ for the one-neutron pickup reaction \su{9}Be(\su{36}S,\su{34}Al) (77.5 MeV$/$nucleon).  The red squares are the experimental data points and the blue dashed line is the previous simulation result, both from Ref.\ \cite{Tur}.  The grey band represents the range of the present simulation results within a $1\sigma$ distribution.} 
\label{fig:34Alband}
\end{figure}

The simulation results for the one nucleon pickup processes are shown in Fig.\ \ref{fig:37Kband} and \ref{fig:34Alband}.  Angular distributions, out-of-plane scattering, and \g-ray deorientation were implemented in the nucleon pickup process as well. In the momentum distribution calculations discussed previously, Souliotis \et \cite{Sou} used the ``typical'' Fermi momentum of 230 MeV$/$c.  In order for the simulation to encompass a range of targets, the Fermi momentum was taken not as 230 MeV$/$c, but was calculated based on data taken from Moniz \et \cite{Mon}.  The Fermi momentum ranges from 170 MeV$/$c for the lightest targets to 260 MeV$/$c for heavier targets.  The results of the simulation for proton pickup are shown in Fig.\ \ref{fig:37Kband}, calculated for the one proton pickup reaction, \su{9}Be(\su{36}Ar,\su{37}K)X, first observed by Groh \et \cite{Gro}.  The momentum matching conditions \cite{Bri} for simple surface-to-surface pickup are best met for the two data points on the high momentum side of the yield distribution, where the simulation agrees with the data. On the low momentum side of the peak of the yield curve, the picked-up nucleon has a momentum less than the Fermi momentum, and the momentum matching conditions for direct pickup are poorly satisfied.  More complex transfer mechanisms are therefore required to describe the polarization on the low momentum side \cite{Gro}. 

The simulation code was also used to model the neutron pickup data obtained in Ref.\ \cite{Tur}, as shown in Fig.\ \ref{fig:34Alband}.  Again, no scaling factor was required to achieve a better quantitative agreement with experiment.

\section{Summary}
A statistical Monte Carlo code for nuclear spin polarization has been modified to include corrections to the kinematical model that aim to improve quantitative agreement with experiment.  The positions of removed nucleons are assigned in a Monte Carlo fashion, and projectiles are allowed to interact on either side of the target.  The real part of the optical potential, $\msub{V}{0}$, is obtained with a folding model calculation and then used in the calculation of the mean deflection angle, \tdef.  Fragments are permitted to scatter to a distribution of deflection angles rather than a single mean angle, and the out of reaction plane acceptance has been taken into account.  The process of de-orientation due to \g-ray relaxation is also included.  The angular distribution implementation reduces the calculated polarization magnitude by about 10\% and the out-of-plane acceptance reduces the polarization magnitude by about 40\%.  The calculated polarization magnitude is corrected by about 50\% due to \g-ray de-orientation.  These corrections account for the 0.25 scaling factor needed by both Ref.\ \cite{Oku} and Ref.\ \cite{Tur} in their simulations. 

It should be noted that the corrections are approximations.  A primitive function is used to calculate the distribution of deflections angles, and no attempt was made to rigorously reproduce the scattering angles observed in fragmentation reactions.  The out-of-plane scattering correction is implemented as a constant multiplicative factor based on the expected acceptances of typical fragment separators.  The deorientation correction due to \g-ray relaxation is calculated in a statistical manner considering average excitation energy and entry spin.  Even within these limitations, the magnitude with which the corrections are shown to improve the results indicate that these are important contributions to consider in the simulation of polarization in intermediate-energy reactions.

\section{Acknowledgements}
We acknowledge the assistance of K. Asahi and W.D. Schmidt-Ott for providing the initial Monte Carlo simulation codes for spin polarization. We thank D.J. Morrissey and P.G. Hansen for helpful discussions on spin polarization.  The work was supported in part by the National Science Foundations grants PHY-06-06007 and PHY-99-83810.

\end{document}